\documentclass[a4paper,twoside,11pt,english,fleqn]{article}

\usepackage[intlimits]{amsmath}
\usepackage{amssymb,amsthm}
\usepackage{graphicx}
\usepackage[dvips]{epsfig}
\usepackage{babel}
\usepackage[T1]{fontenc}
\usepackage{fancyhdr,stmaryrd,color,framed,rotating}
\usepackage{times,euler,euscript,eufrak}
\usepackage{url,calc,xspace}
\usepackage[modulo]{lineno}

\DeclareMathAlphabet{\mathscr}{T1}{pzc}{m}{it}

  	\newtheorem{theorem}{Theorem}[section]
  	
  	\newtheorem{lemma}[theorem]{Lemma}
  	\newtheorem{proposition}[theorem]{Proposition}
\theoremstyle{definition}
	\newtheorem{definition}[theorem]{Definition}
	
	\newtheorem{remark}[theorem]{Remark}
	\newtheorem{example}[theorem]{Example}

\fancypagestyle{plain}{
  \fancyhf{}\fancyfoot[LC,RC]{}
  \fancyfoot[LE,RO]{$\thepage$}
  
  }

\DeclareMathOperator{\id}{Id}
\renewcommand{\phi}{\varphi}
\renewcommand{\epsilon}{\varepsilon}
\newcommand{\fl}{\to}
\newcommand{\dfl}{\Rightarrow}
\newcommand{\tfl}{\Rrightarrow}

\newcommand{\abs}[1]{\left|#1\right|}
\newcommand{\norm}[1]{\abs{\abs{#1}}}
\newcommand{\tnorm}[1]{\abs{\abs{\abs{#1}}}}

\newcommand{\roundup}[1]{\left\lceil#1\right\rceil}
\newcommand{\rounddown}[1]{\left\lfloor#1\right\rfloor}

\newcommand{\Nb}{\mathbb{N}}

\newcommand{\Mr}{\EuScript{M}}

\renewcommand{\Pr}{\EuScript{P}}
\newcommand{\Qr}{\EuScript{Q}}

\newcommand{\Vr}{\EuScript{V}}
\definecolor{rouge}{rgb}{1,0,0}

\definecolor{bleu}{rgb}{0,0,1}

\newcommand{\figt}[1]{\raisebox{-1mm}{\includegraphics{#1.eps}}}
\newcommand{\gfigt}[1]{\raisebox{-2.5mm}{\includegraphics{#1.eps}}}
\newcommand{\ggfigt}[1]{\raisebox{-3.75mm}{\includegraphics{#1.eps}}}
\newcommand{\ggggfigt}[1]{\raisebox{-6.5mm}{\includegraphics{#1.eps}}}

\newcommand{\figeps}[1]{\raisebox{-1.25mm}{\includegraphics{#1.eps}}}
\newcommand{\figepsind}[1]{\raisebox{-1.25mm}{\includegraphics[scale=0.66]{#1.eps}}}

\newcommand{\Ptime}{\textsc{ptime}\xspace}
\newcommand{\NPtime}{\textsc{nptime}\xspace}

\newcommand{\adressea}{\footnote{guillaume.bonfante@loria.fr.}}
\newcommand{\adresseb}{\footnote{yves.guiraud@loria.fr.}}
\newcommand{\umr}{\footnote{UMR 7503 CNRS -- INPL -- INRIA -- UHP -- Nancy 2.}}
\newcommand{\titre}{Intensional properties of polygraphs}
\newcommand{\auteura}{Guillaume Bonfante\adressea}
\newcommand{\auteurb}{Yves Guiraud\adresseb}
\newcommand{\adresse}{INRIA -- LORIA\umr}
\begin{document}

\thispagestyle{empty}
\begin{center}
\textbf{\LARGE \titre{}}

\vspace{3mm}
\textbf{\large \auteura{} -- \auteurb}

\adresse

January 8, 2006 -- Modified October 2, 2007
\end{center}

\vspace{2mm}
\begin{em}
\hrule height 1.5pt

\medskip







\noindent\textbf{Abstract --} We present polygraphic programs, a subclass of Albert Burroni's polygraphs, as a computational model, showing how these objects can be seen as first-order functional programs. We prove that the model is Turing complete. We use polygraphic interpretations, a termination proof method introduced by the second author, to characterize polygraphic programs that compute in polynomial time. We conclude with a characterization of polynomial time functions and non-deterministic polynomial time functions.
\medskip \hrule height 1.5pt
\end{em}

\section{Introduction}
\label{Section:Introduction}

Polygraphs are special higher-dimensional categories, introduced by Albert Burroni to provide a unified algebraic setting for rewriting~\cite{Burroni93}. For example, any term rewriting system can be translated into a polygraph which has, in case of left-linearity, exactly the same properties of termination and confluence~\cite{Lafont03,Guiraud06jpaa}.

Here, we study how these mathematical objects can be used as a computational model. Informally, computations generated by a polygraph are done by a net of cells which individually behave according to some local transition rules. This model is close to John von Neumann's cellular automata~\cite{vonNeumann66} and Yves Lafont's interaction nets~\cite{Lafont90} with notable differences: while von Neumann's automata are essentially synchronous, interaction nets and polygraphs are asynchronous; polygraphs have a much more rigid geometry than interaction nets: the underlying graphs of the formers are directed acyclic graphs, preventing the "vicious circles" of the latters.

Termgraph rewriting systems provide another model of graphical computation~\cite{Plump99}: it is an extension of term rewriting with an additional operation, sharing, that allows for a more correct representation of actual computation. The translation of terms into polygraphs is close to the one into termgraphs and they seem to have the same properties, as suggested by the first results in~\cite{Guiraud07}. For example, let us consider the following term rewriting rule, used to compute the multiplication on natural numbers: $\mathtt{mult}(x,\mathtt{succ}(y)) \fl \mathtt{add}(x,\mathtt{mult}(x,y))$. When applied, this rule duplicates the term corresponding to the argument $x$. In termgraph rewriting, one is able to share it instead, so that there is no need for extra memory space. This sharing operation can be algebraically formalized as an operation with one input and two outputs, whose semantics is a duplication operation. In polygraphs, one can have many such operations with many outputs, explicitely represented and handled.

This is a key fact in our results on implicit computational complexity: indeed, the interpretations we consider here, called \emph{polygraphic interpretations}~\cite{Guiraud06jpaa,Guiraud07}, can reflect the fact that two outputs of the same operation have some links between them, as we will see with the example of the list splitting function used in "divide and conquer" algorithms. This allows us to give complexity bounds where traditional polynomial interpretations~\cite{Lankford79} cannot with the method described in~\cite{CichonLescanne92,BonfanteCichonMarionTouzet01} or to give better bounds, as indicated here and in~\cite{Guiraud07}. Moreover, the polygraphic interpretations give separated information on the spatial and on the temporal complexities of functions.

Because of space constraints, we only give the main arguments used to prove the results we present here; the complete proofs can be found in a longer paper, together with extended comments and technical details~\cite{BonfanteGuiraud06}. Let us note that the present paper proposes a slightly more general framework than the other one, allowing non-deterministic polygraphic programs.

In section~\ref{Section:Computation} we introduce the notion of polygraphic program in an informal way and give the corresponding semantics we consider; we introduce the leading example of this paper, the polygraphic program computing the "fusion sort" on lists, and we prove that polygraphic programs form a Turing complete model of computation. In section~\ref{Section:Interpretations}, we recall the notion of polygraphic interpretation, give examples, define the notion of simple polygraphic program and prove results on termination of polygraphic programs. Finally, in section~\ref{Section:Complexity}, we give polynomial complexity bounds for simple programs and prove that they characterize the classes \Ptime and \NPtime of functions computable in polynomial time, respectively by a Turing machine and by a non-deterministic Turing machine.

\section{Polygraphs as a computational model}
\label{Section:Computation}

The general definition of polygraph can be found in documents by Albert Burroni, Yves Lafont and François Métayer~\cite{Burroni93,Lafont03,Metayer03,Lafont06,LafontMetayer06}. Here we give a rewriting-minded presentation of a special case of polygraphs, seeing them as rewriting systems on algebraic circuits.

\begin{definition} \label{Definition:Polygraph}
A \emph{monoidal $\mathit{3}$-polygraph} is a composite object consisting of \emph{cells}, \emph{paths} and \emph{compositions} organized into \emph{dimensions}.

\emph{Dimension $\mathit{1}$} contains elementary sorts called \emph{$\mathit{1}$-cells} and represented by wires. Their concatenation~$\star_0$ yields product types called \emph{$\mathit{1}$-paths} and pictured as juxtaposed vertical wires. The empty product~$\ast$ is also a $1$-path, represented by the empty diagram.

\emph{Dimension $\mathit{2}$} is made of operations called \emph{$\mathit{2}$-cells}, with a finite number of typed inputs and outputs. They are pictured as circuit gates, with inputs at the top and outputs at the bottom. Using all the $1$-cells and $2$-cells as generators, one builds circuits called \emph{$\mathit{2}$-paths}, using the following two compositions:
\begin{center}\input{2-compositions.pstex_t}\end{center}

\noindent The constructions are considered \emph{modulo} some relations, including topological deformation: one can stretch or contract wires freely, move $2$-cells, provided one does not create crossings or break wires. Each $2$-cell and each $2$-path $f$ has a $1$-path $s_1(f)$ as input, its \emph{$\mathit{1}$-source}, and a $1$-path $t_1(f)$ as output, its \emph{$\mathit{1}$-target}. The compact notation $f:s_1(f)\dfl t_1(f)$ summarizes these facts.

\emph{Dimension $\mathit{3}$} contains rewriting rules called \emph{$\mathit{3}$-cells}. They always transform a $2$-path into another one with the same $1$-source and the same $1$-target. Using all the $1$-cells, $2$-cells and $3$-cells as generators, one can build reductions paths called \emph{$\mathit{3}$-paths}, by application of the following three compositions, defined for~$F$ going from $f$ to $f'$ and~$G$ going from $g$ to $g'$: $F\star_0 G$ goes from $f\star_0 g$ to $f'\star_0 g'$; when $t_1(f)=s_1(g)$, then $F\star_1 G$ goes from $f\star_1g$ to $f'\star_1 g'$; when $f'=g$, then $F\star_2 G$ goes from $f$ to $g'$. These constructions are identified \emph{modulo} some relations, given in~\cite{Guiraud06apal}, where their $3$-dimensional nature was explained. The relations allow one to freely deform the constructions in a reasonable way: in particular, they identify paths that only differ by the order of application of the same $3$-cells on non-overlapping parts of a $2$-path. A $3$-path is \emph{elementary} when it contains exactly one $3$-cell. Each $3$-cell and each $3$-path $F$ has a $2$-path $s_2(F)$ as left-hand side, its \emph{$\mathit{2}$-source}, and a $2$-path $t_2(F)$ as right-hand side, its \emph{$\mathit{2}$-target}. The notation $F:s_2(F)\tfl t_2(F)$ stands for these facts.

For monoidal $3$-polygraphs, rewriting notions are defined in a similar way as for term rewriting systems, with terms replaced by $2$-paths, reduction steps by elementary $3$-paths and reduction paths by $3$-paths~\cite{Guiraud06jpaa}. Hence, a \emph{normal form} in a polygraph $\Pr$ is a $2$-path~$f$ which is the $2$-source of no elementary $3$-path. The polygraph $\Pr$ \emph{terminates} when it does not contain infinite families $(F_n)_{n\in\Nb}$ of elementary $3$-paths such that $t_2(F_n)=s_2(F_{n+1})$ for all $n$. Other rewriting properties, such as \emph{confluence} or \emph{convergence} are also defined in an intuitive way.
\end{definition}

\begin{remark}
As defined here, the $k$-paths of such a polygraph, equipped with their $k$ compositions, form a $k$-category. More precisely, this is the $k$-category freely generated by all the $i$-cells, for $0\leq i\leq k$~\cite{Burroni93}.
\end{remark}

\begin{definition}\label{Definition:PolygraphicProgram}
A \emph{polygraphic program} is a monoidal $3$-polygraph such that:
\begin{itemize}
\item Its $2$-cells are divided into \emph{structure $\mathit{2}$-cells}, \emph{constructors} and \emph{functions}. The structure $2$-cells consist of one $\figeps{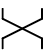}:\xi\star_0\zeta \dfl \zeta\star_0\xi$ for each pair of $1$-cells $(\xi,\zeta)$, plus one $\figeps{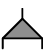}:\xi \dfl \xi\star_0\xi$ and one $\figeps{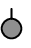}:\xi \dfl \ast$ for each $1$-cell $\xi$. The constructors are $2$-cells such with a $1$-cell as $1$-target. The functions are any $2$-cells.

\item Its $3$-cells are divided between \emph{structure $\mathit{3}$-cells} and \emph{computation $\mathit{3}$-cells}. The structure $3$-cells are given, for every constructor $\figeps{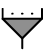}:x\dfl\xi$ and every $1$-cell $\zeta$, by:
\begin{center}\input{3-cellules-structure-2.pstex_t}\end{center}

\noindent The $2$-targets of the $3$-cells of $\Delta_3^2$ use structure $2$-paths built from the structure $2$-cells by using the following structural induction rules:
\begin{center}\input{2-chemins-structure.pstex_t}\end{center}

\noindent The computation $3$-cells are $3$-cells whose $2$-source is of the shape $t\star_1\phi$, with $\phi$ a function $2$-cell and $t$ a $2$-path built only with $1$-cells and constructors. Furthermore, there is a finite constant that bounds the number of structure $2$-cells in the $2$-target of each computation $3$-cell.

\item For the present study, we assume that there exists a procedure to perform each step of computation: more formally, for every $3$-path $F:f\fl g$ containing exactly one $3$-cell, the map giving $g$ from the pair $(f,F)$ is computable in polynomial time.
\end{itemize}
\end{definition}

\vfill\pagebreak
\begin{example}\label{Example:Arithmetics}
We consider the following polygraphic program with one $1$-cell, two constructors $\figeps{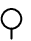}$ and~$\figeps{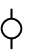}$, two functions $\figeps{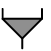}$ and $\figeps{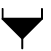}$ and four computation $3$-cells (we do not give the structure cells):
\begin{center}\input{3-cellules-polynomes.pstex_t}\end{center}

\noindent With the constructors, one can represent the natural number $n$, using $\figeps{cons-0}$ for $0$ and $\figeps{cons-1}$ for the successor operation, yielding a $2$-path $t_n$ with zero input and one output. Furthermore, one can check that this polygraph is convergent and that, given $t_m$ and $t_n$, the normal form of $(t_m\star_0 t_n)\star_1\figeps{fonction-2-b}$ is $t_{m+n}$, while the one of $(t_m\star_0 t_n)\star_1\figeps{fonction-2}$ is $t_{mn}$.

Hence this polygraphic program computes the addition and the multiplication on natural numbers: the $1$-cells are the data types, the $2$-paths $\xi\dfl\ast$ built only from constructors are the values, while the result of the application of a function $\figeps{phi}$ with $n$ inputs to well-typed values $(t_1,\dots,t_n)$ is the normal form of the $2$-path $(t_1\star_0\cdots\star_0 t_n)\star_1\figeps{phi}$. This semantical interpretation is formalized thereafter.
\end{example}

\begin{definition}[Semantics] \label{Definition:Semantics}
Let us fix a polygraphic program $\Pr$. If $\xi$ is a $1$-cell, a \emph{term of type $\mathit{\xi}$} is a $2$-path built only with constructors and with $\xi$ as $1$-target. A \emph{value} or \emph{closed term} is a term with no input. The set of values with type $\xi$ is denoted by~$\Vr(\xi)$. The \emph{domain of computation} of $\Pr$ is the multi-sorted algebra made of the family of all the sets $\Vr(\xi)$ equipped with the operations given, for each constructor $\gamma:\xi_1\star_0\cdots\star_0\xi_n\dfl \xi$, by the map still denoted by $\gamma$:
\begin{align*}
\gamma \: : \:
\Vr(\xi_1)\times\cdots\times\Vr(\xi_n) &\fl \Vr(\xi) \\
(t_1,\dots, t_n) &\mapsto (t_1\star_0\cdots\star_0 t_n) \star_1 \gamma.
\end{align*}

\noindent Let us consider a function $f$ from $\Vr(\xi_1)\times\cdots\times\Vr(\xi_m)$ to $\Vr(\zeta_1) \times \cdots \times \Vr(\zeta_n)$. There are two cases whether $\Pr$ is confluent or not.

If $\Pr$ is confluent, we say that $\Pr$ \emph{computes} $f$ if there exists a $2$-path, still denoted by $f$, from $\xi_1\star_0\cdots\star_0\xi_m$ to $\zeta_1\star_0\cdots\star_0\zeta_n$, such that, for every family $(t_1,\dots,t_m)$ of values in $\Vr(\xi_1)\times\cdots\times\Vr(\xi_m)$, the $2$-path $(t_1\star_0\dots\star_0 t_m)\star_1 f$ normalizes into the family $f(t_1,\dots,t_m)$ of values in $\Vr(\zeta_1) \times \cdots \times \Vr(\zeta_n)$. Let us note that the normal form is unique in that case.

If $\Pr$ is not confluent, after Gurevich and Gr\"adel \cite{GradelGurevich92}, we say that it computes $f$ if, for all values $(t_1,\dots,t_m)$ in $\Vr(\xi_1)\times\cdots\times\Vr(\xi_m)$, the following holds, where the considered order is the lexicographic order:
$$
f(t_1,\dots,t_m) \:=\: \max \big\{ \:\text{normal forms of } (t_1\star_0\dots\star_0 t_m)\star_1 f \:\big\}.
$$
\end{definition}

\begin{example}\label{Example:FusionSort}
Let us consider a polygraphic program that computes, among other functions, the \emph{fusion sort} function on lists of natural numbers. It has two $1$-cells, $\mathtt{nat}$ for natural numbers and $\mathtt{list}$ for lists of natural numbers. Its other cells, apart from structure ones are:
\begin{itemize}
\item Constructors: one $\raisebox{-1mm}{\begin{picture}(0,0)%
\includegraphics{n.pstex}%
\end{picture}%
\setlength{\unitlength}{4144sp}%
\begingroup\makeatletter\ifx\SetFigFont\undefined%
\gdef\SetFigFont#1#2#3#4#5{%
  \reset@font\fontsize{#1}{#2pt}%
  \fontfamily{#3}\fontseries{#4}\fontshape{#5}%
  \selectfont}%
\fi\endgroup%
\begin{picture}(139,174)(887,1097)
\put(957,1188){\makebox(0,0)[b]{\smash{{\SetFigFont{6}{7.2}{\rmdefault}{\mddefault}{\updefault}{\color[rgb]{0,0,0}$n$}%
}}}}
\end{picture}%
}:\ast\dfl\mathtt{nat}$ for each natural number $n$, plus $\figt{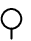}:\ast\dfl\mathtt{list}$ for the empty list and $\figt{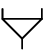}:\mathtt{nat}\star_0\mathtt{list}\dfl\mathtt{list}$ for the list constructor.

\item Functions: the main $\figt{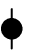}:\mathtt{list}\dfl\mathtt{list}$ for fusion sort, together with $\figt{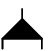}:\mathtt{list}\dfl\mathtt{list}\star_0\mathtt{list}$ for splitting lists and $\figt{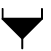}:\mathtt{list}\star_0\mathtt{list}\dfl\mathtt{list}$ for merging them.

\vfill\pagebreak
\item Computation $3$-cells:
\begin{center}\input{tri-rapide-sort.pstex_t}\end{center}
\begin{center}\input{tri-rapide-split.pstex_t}\end{center}
\begin{center}\input{tri-rapide-merge.pstex_t}\end{center}
\end{itemize}

\noindent Note that the last two rules for the $\figt{merge}$ function are, in fact, a notation for an infinite family of $3$-cells: there is exactly one of them for each pair $(p,q)$ of natural numbers, depending which one of $p\leq q$ or $p>q$ holds. However, these two conditions are computable (in linear time), preventing super-Turing computations. We have chosen a simplified representation of natural numbers which considers them as being predefined, at the "hardware level", together with their predicate $\leq$. The reason for this choice is to postpone the study of modularity and of the $\mathtt{if}$-$\mathtt{then}$-$\mathtt{else}$ construction to subsequent work.
\end{example}

\begin{theorem}\label{Theorem:TuringComplete}
Polygraphic programs form a Turing-complete model of computation.
\end{theorem}

\begin{proof}
Here we give a sketch of the proof, while the complete one can be found in~\cite{BonfanteGuiraud06}. We could use the fact that we can simulate rewriting systems which are Turing complete, but we give the explicit simulation as a first step to the proof of Theorem~\ref{Theorem:PTIME}. Given a Turing machine, one defines:
\begin{itemize}
\item Constructors: one $\figt{nil}:0\dfl 1$ for the empty word plus one $\raisebox{-1mm}{\begin{picture}(0,0)%
\includegraphics{a.pstex}%
\end{picture}%
\setlength{\unitlength}{4144sp}%
\begingroup\makeatletter\ifx\SetFigFont\undefined%
\gdef\SetFigFont#1#2#3#4#5{%
  \reset@font\fontsize{#1}{#2pt}%
  \fontfamily{#3}\fontseries{#4}\fontshape{#5}%
  \selectfont}%
\fi\endgroup%
\begin{picture}(123,204)(344,557)
\put(406,637){\makebox(0,0)[b]{\smash{{\SetFigFont{6}{7.2}{\rmdefault}{\mddefault}{\updefault}{\color[rgb]{0,0,0}$a$}%
}}}}
\end{picture}%
}:1\dfl 1$ for each letter $a$.
\item Functions: one $\figt{sort}:1\tfl 1$ for the function to be computed plus one $\mathtt{step}_{q,a}=\raisebox{-1mm}{\input{step-q-a.pstex_t}}:2\dfl 1$ for each state $q$ and each letter $a$, including the blank symbol $\sharp$.
\item Computation $3$-cells are given thereafter, the first rule initializing the computation, the four subsequent families replicating the transitions of the Turing machine and the final family starting the computation of the result:
\begin{displaymath}
\raisebox{-1mm}{\input{3-cellules-turing.pstex_t}}
\end{displaymath}
\end{itemize}

\noindent We assume that, at the end of the computation, the result of the Turing machine is the word written on the tape at the right of the head: this is why the last $3$-cell erases the left part.

Let us assume that the machine is in the state $q$, reading the letter $a$, with $w_l$ and $w_r$ the two words respectively written on the left of $a$, from right to left, and on the right of $a$, from left to right. Then, this state of the system is represented by the $2$-path $(w_l\star_0 w_r)\star_1\raisebox{-1.25mm}{\input{step-q-a.pstex_t}}$. Then one checks that each transition step of the Turing machine corresponds to an elementary $3$-path of its polygraphic version.
\end{proof}

\section{Polygraphic interpretations and simple programs}
\label{Section:Interpretations}

Polygraphic interpretations have been introduced to prove termination of $3$-polygraphs~\cite{Guiraud06jpaa}. Here we use a restricted form to get properties on the complexity of polygraphic programs. In order to give some intuition, let us imagine that $2$-paths are electrical circuits, crossed by currents going downwards, from the inputs to the outputs.

A polygraphic interpretation associates to each $2$-cell $\phi$ a "current map" $\phi_*$ and a "heat map" $[\phi]$. The current map gives the intensity of the currents leaving the circuit gate $\phi$, according to the intensity of the incoming currents. The heat map says how much heat is produced by the same gate $\phi$ during this process.

From these maps, one can compute the currents and heat produced by each $2$-path. The original use was to find a polygraphic interpretation such that each reduction step replaces a $2$-path by another one that produces strictly less heat. Here, the current maps will be used as approximations of the size of the computed values, while heat maps will estimate the number of computation steps remaining to reach a result: hence, current maps and heat maps will give bounds respectively on the spatial and on the temporal complexities of a polygraphic program.

\begin{definition}\label{Definition:Interpretation}
A \emph{polygraphic interpretation} of a polygraphic program $\Pr$ consists into a mapping of each $2$-path $f$ with $m$ inputs and $n$ outputs onto two monotone maps $f_* : \Nb^m \fl \Nb^n$ and $[f] : \Nb^m \fl \Nb$, such that the following conditions are satisfied:
\begin{itemize}
\item For every $1$-path $x$ of length $n$, we have $x_*=\id_{\Nb}^n$ and $[x]=0$.
\item For every $2$-paths $f$ and $g$, the following equalitities hold when defined:
\begin{itemize}
\item $(f\star_0 g)_*(\vec{x},\vec{y}) = (f_*(\vec{x}),g_*(\vec{y}))\:$ and $\:[f\star_0 g](\vec{x},\vec{y}) = [f](\vec{x}) + [g](\vec{y})\:$;
\item $(f\star_1 g)_*(\vec{x}) = g_*(f_*(\vec{x}))\:$ and $\:[f\star_1 g](\vec{x}) = [f](\vec{x}) + [g](f_*(\vec{x}))$.
\end{itemize}
\end{itemize}

\noindent Given an interpretation and a $2$-cell $\phi$, we denote by $\phi^j_*$ the $j^{\text{th}}$ component of the map $\phi_*$. An interpretation of $\Pr$ generates a binary relation denoted by $\succ$: it is defined, on $2$-paths $f$ and $g$ with the same $2$-source and the same $2$-target, by $f\succ g$ when the two inequalities $f_*(\vec{i}) \geq g_*(\vec{i})$ and $[f](\vec{i}) > [g](\vec{i})$ hold for every possible family $\vec{i}$ of natural numbers. An interpretation is \emph{compatible} with a $3$-cell $\alpha$ when $s_2(\alpha)\succ t_2(\alpha)$ and \emph{weakly compatible} with $\alpha$ if $s_2(\alpha)\succeq t_2(\alpha)$.
\end{definition}

\noindent It was proved in~\cite{Guiraud06jpaa} that an interpretation is entirely determined by its values on the $2$-cells of the polygraph, that the binary relation $\succ$ is a terminating strict order and that context are strictly monotone with respect to it. These are the main steps towards the following result.

\begin{theorem}[\cite{Guiraud06jpaa}]\label{Theorem:Termination}
If a polygraphic program admits an interpretation which is compatible with all of its $3$-cells, then it terminates.
\end{theorem}

\begin{example}\label{Example:StructureHeat}
Let us assume that we have a current map $(\cdot)_*$ on a polygraphic program such that the following conditions hold:
\begin{itemize}
\item If $\figeps{phi}$ is a constructor with $n$ inputs, then $\figeps{phi}_*(i_1,\dots,i_n)>i_1+\cdots+i_n$.
\item One structure $2$-cells, we have $\figeps{tau}_*(i,j)=(j,i)$ and $\figeps{delta}_*(i)=(i,i)$.
\end{itemize}

\noindent We define a heat map $[\cdot]_S$ as follows:
\begin{itemize}
\item If $\figeps{phi}$ is a constructor or a function, then $\left[\figeps{phi}\right]_S=0$.
\item On structure $2$-cells, we have $\left[\figeps{tau}\right]_S(i,j)=ij$, $\left[\figeps{delta}\right]_S(i)=i^2$ and $\left[\figeps{epsilon}\right]_S(i)=i$.
\end{itemize}

\noindent It is proved in~\cite{BonfanteGuiraud06} that these values generate a polygraphic interpretation compatible with the structure $3$-cells. Hence theorem~\ref{Theorem:Termination} tells us that a polygraphic program without computation $3$-cell terminates.
\end{example}

\begin{definition}\label{Definition:StructureHeat}
Given a current map $(\cdot)_*$ on a polygraphic program that satisfies the conditions of example~\ref{Example:StructureHeat}, the heat map $[\cdot]_S$ is called \emph{structure heat} generated by $(\cdot)_*$.
\end{definition}

\begin{definition}\label{Definition:Simple}
We denote by $\Nb[x_1,\cdots,x_n]$ the algebra of polynomials in $n$ variables and coefficients in~$\Nb$. Let $\Pr$ be a polygraphic program. A polygraphic interpretation is \emph{simple} when the following conditions are met:
\begin{itemize}
\item For any $2$-cell $\phi$ with $m$ inputs and $n$ outputs, the maps $\sum_{j=1}^n \phi_*^j$ and $[\phi]$ are in $\Nb[x_1,\dots,x_m]$.
\item If $\gamma$ is a constructor with $n$ inputs, then $\gamma_* = \sum_{i=1}^m x_i + a_{\gamma}$, with $a_{\gamma}>0$, and $[\gamma] = 0$. Moreover, we assume that there exists a $a\in\Nb^*$ bounding all the $a_{\gamma}$'s.
\item On structure $2$-cells, one has $\figeps{tau}(i,j)=(j,i)$ and $\figt{delta}(i)=(i,i)$. Moreover,
structure cells produce no heat: $\left[ \figeps{delta} \right] (i) = 0 , \left[ \figeps{tau} \right] (i,j) = 0 , \left[ \figeps{epsilon} \right] (i) = 0$.
\item For every function $\phi$ with $m$ inputs and $n$ outputs and for every family $(i_1,\dots,i_m)$ of natural numbers, we have $\sum_{j=1}^n \phi^j_*(i_1,\dots,i_m) \:\geq\: i_1+\cdots+i_m$.
\end{itemize}

\noindent A polygraphic program is called \emph{simple} when the $2$-targets of its computation $3$-cells contain at most~$K$ structure $2$-cells, for some fixed $K$, and when it admits a simple polygraphic interpretation which is compatible with all of its computation $3$-cells.
\end{definition}

\begin{theorem}\label{Theorem:Termination2}
A simple polygraphic program terminates.
\end{theorem}

\begin{proof}
Let $\Pr$ be a simple polygraphic program and let $(\cdot)_*$ and $[\cdot]$ be the current and heat maps of a simple interpretation, compatible with all the computation $3$-cells of $\Pr$. It is a direct computation to check that such an interpretation is weakly compatible with the structure $3$-cells of $\Pr$. Hence, we deduce that $\Pr$ terminates if and only if the polygraphic program $\Qr$ does, where $\Qr$ is built from $\Pr$ by removal of the computation $3$-cells. The map $(\cdot)_*$ also satisfies the conditions to generate a structure heat map $[\cdot]_S$ proving the termination of $\Qr$.
\end{proof}

\begin{example}\label{Example:FusionSortInterpretation}
Let us prove that the polygraphic program of example~\ref{Example:FusionSort} is simple. Let us consider the interpretation generated by these values:
\begin{itemize}
\item $\raisebox{-1.25mm}{\begin{picture}(0,0)%
\includegraphics{n.pstex}%
\end{picture}%
\setlength{\unitlength}{4144sp}%
\begingroup\makeatletter\ifx\SetFigFont\undefined%
\gdef\SetFigFont#1#2#3#4#5{%
  \reset@font\fontsize{#1}{#2pt}%
  \fontfamily{#3}\fontseries{#4}\fontshape{#5}%
  \selectfont}%
\fi\endgroup%
\begin{picture}(139,174)(887,1097)
\put(957,1188){\makebox(0,0)[b]{\smash{{\SetFigFont{6}{7.2}{\rmdefault}{\mddefault}{\updefault}{\color[rgb]{0,0,0}$n$}%
}}}}
\end{picture}%
}_*=1$, $\quad\figeps{nil}_*=1$, $\quad\figeps{cons}_*(i,j)=i+j+1$;
\item $\figeps{sort}_*(i)=i$, $\quad\figeps{split}_*(i) = (\roundup{i/2} ,\rounddown{i/2})$, $\quad\figeps{merge}_*(i,j)=i+j$;
\item $\left[ \figeps{sort} \right] (i) = 2i^2$, $\quad\left[ \figeps{split} \right] (i) = i$, $\quad\left[ \figeps{merge} \right](i,j) = i+j$.
\end{itemize}

\noindent We have used the notations $\roundup{\cdot}$ and $\rounddown{\cdot}$ for the rounding functions, respectively by excess and by default. This interpretation meets the conditions of definition~\ref{Definition:Simple} and, thus, is simple. Now, one has to check that it is compatible with all the computation $3$-cells: we give some of the computations for the last $3$-cell of the function $\figeps{sort}$. Let us start with $(\cdot)*$. On one hand:
$$
\left( \ggfigt{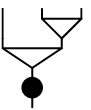}  \right)_* (i,j,k)
\:=\: \left ( \gfigt{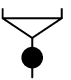} \right)_* \left( i,\figeps{cons}_*(j,k)\right)
\:=\: \figeps{sort}_* \circ \figeps{cons}_* \left( i, \figeps{cons}_*(j,k)\right) \\
\:=\: i+j+k+2.
$$

\noindent And, on the other hand:
\begin{displaymath}
\left( \ggggfigt{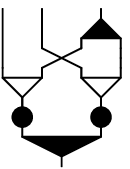} \right)_* (i,j,k)
\: = \: i+j+\roundup{k/2}+\rounddown{k/2}+2
\: = \: i+j+k+2.
\end{displaymath}

\noindent Now, let us consider $[\cdot]$. For the $2$-source of the $3$-cell, one gets:
\begin{displaymath}
\left[ \ggfigt{2-source-sort-1} \right] (i,j,k)
\:=\: \left[ \figeps{sort} \right] (i+j+k+2)
\:=\: 2\cdot(i+j+k+2)^2.
\end{displaymath}

\noindent And, for its $2$-target, $\left[ \ggggfigt{2-but-sort} \right] (i,j,k)$ is equal to:
\begin{align*}
&\!\!\!\!\!\! \left[ \figeps{split} \right] (k)
+ \left[ \figeps{sort} \right] \left( i + \roundup{k/2} +1 \right)
+ \left[ \figeps{sort} \right] \left( j+ \rounddown{k/2} +1 \right)
+ \left[ \figeps{merge} \right] \left ( i + \roundup{k/2} +1, j+ \rounddown{k/2} +1 \right) \\
=&\: 2\cdot\left( i + \roundup{k/2} +1 \right)^2
+ 2\cdot\left( j+ \rounddown{k/2} +1 \right)^2
+ i +j + 2k + 2.
\end{align*}

\noindent We conclude, for example, by considering two cases, depending on the parity of $k$.
\end{example}

\begin{example}
For the polygraphic program of example~\ref{Example:Arithmetics}, the following values generate a simple interpretation which is compatible with the four computation $3$-cells:
\begin{itemize}
\item $\figeps{cons-0}_*=1$, $\quad\figeps{cons-1}_*(i)=i+1$, $\quad\figeps{delta}_*(i)=(i,i)$, $\quad\figeps{fonction-2-b}_*(i,j)=i+j$,  $\quad\figeps{fonction-2}_*(i,j)=ij$;
\item $\left[\figeps{cons-0}\right] = \left[\figeps{cons-1}\right](i) = \left[\figeps{delta}\right](i) = \left[\figeps{epsilon}\right](i) = 0$, $\quad\left[\figeps{fonction-2-b}\right](i,j)=i$, $\quad\left[\figeps{fonction-2}\right](i,j)=(i+1)j$.
\end{itemize}
\end{example}

\section{Complexity of simple programs}
\label{Section:Complexity}

\begin{definition}\label{Definition:Sizes}
Let $\Pr$ be a polygraphic program. If $f$ is a $2$-path of $\Pr$, we denote by $\norm{f}$ the number of $2$-cells $f$ is made of. If $F$ is a $3$-path of $\Pr$, we denote by $\tnorm{F}$ the number of $3$-cells $F$ is made of.
\end{definition}

\noindent Let $\Pr$ be a simple program with a fixed interpretation made of $(\cdot)_*$ and $[\cdot]$. We want to prove that $(\cdot)_*$ is a good estimation of the size of values computed by $\Pr$, given by $\norm{\cdot}$, while $[\cdot]$ is one for the size of the computations, given by $\tnorm{\cdot}$. Once again, the complete proofs are in~\cite{BonfanteGuiraud06}. We recall that, by assumption, there exists a $a>0$ that bounds each $a_{\gamma}=\gamma_*(0,\dots,0)$, for $\gamma$ a constructor. By induction on the size of values, we prove that $(\cdot)_*$ is an estimation of the size of values:

\begin{lemma}\label{Lemma:SizeValues}
For every value $t$, the inequalities $\norm{t} \leq t_* \leq a\norm{t}$ hold in $\Nb$.
\end{lemma}

\noindent Using the properties of the polygraphic interpretation we consider and lemma~\ref{Lemma:SizeValues}, we prove that the size of intermediate and of final values are bounded by a polynomial in the size of the initial values:

\begin{proposition}\label{Proposition:SizeResult}
Let $\phi$ be a function with $m$ inputs and $n$ outputs. Let $P_{\phi}$ be the polynomial in $\Nb[x_1,\dots,x_m]$ defined by $P_{\phi}=\sum_{j=1}^n \phi_*^j(ax_1,\dots,ax_m)$. Let $t$ be a family of values of type $s_1(\phi)$ and let us assume that $t\star_1\phi$ reduces into a $2$-path of the shape $u\star_1 c$, where $u$ has $p$ outputs. Then the inequality $\sum_{j=1}^p u^j_* \leq P_{\phi}(\norm{t^{\scriptscriptstyle 1}},\dots,\norm{t^m})$ holds. In particular, if $u=\phi(t)$, the inequality $\norm{\phi(t)} \leq P_{\phi}(\norm{t^{\scriptscriptstyle 1}},\dots,\norm{t^m})$ holds.
\end{proposition}

\begin{example}
If one computes these polynomials for the simple polygraphic program of example~\ref{Example:FusionSort}, one sees that, for any list $t$, the sorted list $\figeps{sort}(t)$ and all the intermediate values computed to reach the result have their sizes bounded by the size of $t$:
\begin{center}
$P_{\figepsind{sort}}(x) \:=\: \figeps{sort}_*(1\cdot x) \:=\: x$, $\qquad\qquad P_{\figepsind{merge}}(x,y) \:=\: \figeps{merge}_*(1\cdot x,1\cdot y) \:=\: x+y$, \\

\smallskip
$P_{\figepsind{split}}(x) \:=\: \figeps{split}^1_*(1\cdot x) + \figeps{split}^2_*(1\cdot x)
\:=\: \roundup{x/2} +\rounddown{x/2} \:=\: x$.
\end{center}

\smallskip
\noindent For the polygraphic program of example~\ref{Example:Arithmetics}, one gets $P_{\figepsind{fonction-2-b}}(x,y)=x+y$ and $P_{\figepsind{fonction-2}}(x,y)=xy$. Hence, the current maps give us information on the spatial complexity of the computation, separated from the length of computations.
\end{example}

\noindent Now we interest ourselves into polynomial bounds for the length of computations. We start by a technical lemma, which proves that, during a computation, the potential structure heat increase due to the application of a computation $3$-cell is polynomially bounded by the size of the arguments. We recall that, by assumption, each computation $3$-cell contains at most $K$ structure $2$-cells.

\begin{lemma}\label{Lemma:SizeManagement}
Let $\phi$ be a function with $m$ inputs. We denote by $S_{\phi}$ the polynomial $K\cdot P_{\phi}^2$. Let $t$ be a family of values of type $s_1(\phi)$, let $f$ and $g$ be $2$-paths such that $t\star_1\phi$ reduces into $f$ which itself reduces into $g$ by application of a computation rule~$\alpha$. Then the following inequality holds:
$$
[f]_S + S_{\phi}(\norm{t^{\scriptscriptstyle 1}},\dots,\norm{t^m}) \geq [g]_S.
$$
\end{lemma}

\begin{proof}
The complete, technical proof is in~\cite{BonfanteGuiraud06}. Here we recall the main reasoning steps. We denote by $\alpha:a\tfl b$ the computation $3$-cell used to reduce $f$ into $g$. We decompose $f$ and $g$ to make $a$ and $b$ appear and use the properties of current and heat maps to conclude that there exist natural numbers $i_1$, $\dots$, $i_m$ such that the inequality $[f]_S + [b]_S(i_1,\dots,i_m) \: \geq \: [g]_S$ holds. Then we prove that $[b]_S(i_1,\dots,i_m)$ is polynomially bounded by the size of $t$. By definition of the structure heat, $[b]_S(i_1,\dots,i_m)$ is the sum of all the structure heats produced by the structure $2$-cells $b$ is made of. Then we use proposition~\ref{Proposition:SizeResult} to prove that the current incoming in each input of each structure $2$-cell of $b$ is bounded by $P_{\phi}(\norm{t^{\scriptscriptstyle 1}},\dots,\norm{t^m})$. Then, by definition of $[\cdot]_S$ on structure $2$-cells, we conclude that the structure heat produced by each one is at most $P^2_{\phi}(\norm{t^{\scriptscriptstyle 1}},\dots,\norm{t^m})$. Finally, we use the fact that $b$ is the $2$-target of a computation $3$-cell to deduce that there is at most $K$ structure $2$-cells in $b$.
\end{proof}

\begin{example}
For the polygraphic program of example~\ref{Example:FusionSort} we have $K=1$,  $S_{\figepsind{sort}} (x) = x^2$, $S_{\figepsind{split}} (x) = x^2$ and $S_{\figepsind{merge}} (x,y) = (x+y)^2$. For the one of example~\ref{Example:Arithmetics}, we have $K=1$, $S_{\figepsind{fonction-2}}(x,y)=(x+y)^2$ and $S_{\figepsind{fonction-2-b}}(x,y)=x^2y^2$.
\end{example}

\noindent Now let us prove that the length of a computation is polynomially bounded by the size of the arguments.

\begin{proposition}\label{Proposition:SizeReductions}
Let $\phi$ be a function with $m$ inputs. We define the following polynomials:
$$
Q_{\phi}(x_1,\dots,x_m) \:=\: [\phi](ax_1,\dots,ax_m) \qquad\text{and}\qquad R_{\phi} \:=\: Q_{\phi}\cdot (1 + S_{\phi}).
$$

\noindent Let $t$ be a family of values of type $s_1(\phi)$, let $F$ be a $3$-path with $2$-source $t\star_1\phi$, made of $k$ computation $3$-cells and $l$ structure $3$-cells. Then the following inequalities hold:
$$
k \leq Q_{\phi}(\norm{t^{\scriptscriptstyle 1}}, \dots, \norm{t^m} ))
\qquad\text{and}\qquad
l \leq Q_{\phi}(\norm{t^{\scriptscriptstyle 1}}, \dots, \norm{t^m} )) \cdot S_{\phi}(\norm{t^{\scriptscriptstyle 1}}, \dots, \norm{t^m} )).
$$

\noindent As a consequence, $\tnorm{F} \leq R_{\phi}(\norm{t^{\scriptscriptstyle 1}},\dots,\norm{t^m})$ holds.
\end{proposition}

\begin{proof}
We decompose $F$ into a $\star_2$-composite of elementary computation $3$-paths followed by structure $3$-paths. Using the fact that the heat map we consider is strictly decreasing on computation $3$-cells and weakly decreasing on structure $3$-cells, we deduce that $[t\star_1\phi]$ is minored by $k$. We use the properties of~$[\cdot]$ and lemma~\ref{Lemma:SizeValues} to get the bound we seek on $k$. Then, we apply proposition~\ref{Proposition:SizeReductions} to each of the structure $3$-paths we have isolated. We sum up the resulting inequalities and use the facts that $[t\star_1\phi]_S=0$ and $[t_2(F)]_S\geq 0$ to get $k\cdot S_{\phi}(\norm{t^{\scriptscriptstyle 1}},\dots,\norm{t^m}) \geq l$. We deduce the inequality on $l$ from this one and the one on $k$. We conclude by using the equality $\tnorm{F}=k+l$.
\end{proof}

\begin{example}
For the functions of example~\ref{Example:FusionSort}, we have $Q_{\figepsind{sort}} (x) = 2x^2$, $Q_{\figepsind{split}} (x) = x$ and $Q_{\figepsind{merge}} (x,y) = x+y$. For example, let us fix a list $t$. The polynomial $Q_{\figepsind{sort}}$ tells us that, during the computation of the sorted list~$\figt{sort}(t)$, there will be at most $\norm{t}$ applications of a computation $3$-cell. The polynomial $R_{\figepsind{sort}}$ guarantees that there is no more than $\norm{t}^2(1+\norm{t}^2)$ applications of rules. On the examples we have considered, the polynomial $Q_{\phi}$ gives a bound that is close to known ones but the polynomial $R_{\phi}$ gives a very overestimated bound. To get a better estimation, we will have to work on the structure heat increase bound $S_{\phi}$.
\end{example}

\begin{theorem}\label{Theorem:PTIME}
Functions computed by simple confluent polygraphic programs are exactly \Ptime functions.
\end{theorem}

\begin{proof}
We start by proving that functions computed by simple polygraphic programs are in \Ptime. Proposition~\ref{Proposition:SizeReductions} tells us that the length of any computation in such a polygraph are polynomially bounded by the size of the arguments. Furthermore, each step of computation can be done in polynomial time with respect to the size of the current $2$-path: we find a redex in a directed acyclic graph with decorations then replace it by the corresponding reduce and both operations can be done in polynomial time.

Now let us prove that any \Ptime function can be computed by a simple polygraphic program. The first step is to translate a Turing machine equipped with a clock into a polygraphic program. We fix a function $f$ in \Ptime, a Turing machine $\Mr$ that computes $f$ and a polynomial $P$ that bounds the length of the computation. We consider a copy of the polygraphic program of example~\ref{Example:Arithmetics} which computes addition and multiplication of natural numbers, with its $1$-source denoted by $\mathtt{nat}$. Let us note that this polygraphic program computes any polynomials, including $P$. Then we extend it with a variant of the polygraphic Turing machine of section~\ref{Section:Computation}: it is made of a $1$-cell $\mathtt{mon}$; its constructors are the empty word $\figeps{nil} : \mathtt{mon} \dfl \mathtt{mon}$, plus one cell $\raisebox{-1.25mm}{\begin{picture}(0,0)%
\includegraphics{a.pstex}%
\end{picture}%
\setlength{\unitlength}{4144sp}%
\begingroup\makeatletter\ifx\SetFigFont\undefined%
\gdef\SetFigFont#1#2#3#4#5{%
  \reset@font\fontsize{#1}{#2pt}%
  \fontfamily{#3}\fontseries{#4}\fontshape{#5}%
  \selectfont}%
\fi\endgroup%
\begin{picture}(123,204)(344,557)
\put(406,637){\makebox(0,0)[b]{\smash{{\SetFigFont{6}{7.2}{\rmdefault}{\mddefault}{\updefault}{\color[rgb]{0,0,0}$a$}%
}}}}
\end{picture}%
} : \mathtt{mon} \dfl \mathtt{mon}$ for each letter of the alphabet of $\Mr$; its functions are the main $\figeps{sort}:\mathtt{mon}\dfl\mathtt{mon}$ for $f$, plus a size function $\figeps{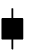}:\mathtt{mon}\dfl\mathtt{nat}$, plus a modified cell $\raisebox{-1.25mm}{\input{clock-step.pstex_t}} : \mathtt{nat}\star_0\mathtt{mon}\star_0\mathtt{mon} \dfl \mathtt{mon}$ for each state $q$ of $\Mr$ and each letter $a$ in the alphabet of $\Mr$, including the blank symbol $\sharp$; its computation $3$-cells are:
\begin{displaymath}
\raisebox{-1.25mm}{\input{3-cellules-turing-clock.pstex_t}}
\end{displaymath}

\noindent Then, one checks that this polygraphic program mimics the transition of the original Turing machine $\Mr$ and, thus, computes $f$. We conclude by checking that the following polygraphic interpretation, extending the one already built on natural numbers, is simple and compatible with each computation $3$-cell:
\begin{itemize}
\item $\figeps{nil}_*=1$, $\:\raisebox{-1.25mm}{\begin{picture}(0,0)%
\includegraphics{a.pstex}%
\end{picture}%
\setlength{\unitlength}{4144sp}%
\begingroup\makeatletter\ifx\SetFigFont\undefined%
\gdef\SetFigFont#1#2#3#4#5{%
  \reset@font\fontsize{#1}{#2pt}%
  \fontfamily{#3}\fontseries{#4}\fontshape{#5}%
  \selectfont}%
\fi\endgroup%
\begin{picture}(123,204)(344,557)
\put(406,637){\makebox(0,0)[b]{\smash{{\SetFigFont{6}{7.2}{\rmdefault}{\mddefault}{\updefault}{\color[rgb]{0,0,0}$a$}%
}}}}
\end{picture}%
}_*(i)=i+1$, $\:\figeps{size}_*(i)=i$, $\:\raisebox{-1.25mm}{\input{clock-step.pstex_t}}_*(i,j,k)=i+j+k$, $\:\figeps{sort}_*(i) \: = \: P_*(i) + i + 1$.
\item $\left[\figeps{size}\right](i) = i$, $\:\left[\raisebox{-1.25mm}{\input{clock-step.pstex_t}}\right](i,j,k)=i$, $\:\left[\figeps{sort}\right](i)=[P](i)+P_*(i)+i+1$.
\end{itemize}
\end{proof}

\noindent Actually, as the simulation is done step by step, another theorem follows, still taking the definition of \NPtime functions to be the one of Gr\"adel and Gurevich~\cite{GradelGurevich92}:

\begin{theorem}\label{Theorem:NPTIME}
Functions computed by simple non-confluent polygraphic programs are exactly \NPtime functions.
\end{theorem}

\bibliographystyle{amsplain}
\begin{small}
\bibliography{bibliographie}
\end{small}
\end{document}